\documentclass[aip,rsi,reprint]{revtex4-1} 
\usepackage{graphicx,color,amsmath,amssymb,chngpage,mathtools,subfigure}

\definecolor{myorange}{rgb}{1,.5,0}
\definecolor{mygreen}{rgb}{0,0.5,0}
\definecolor{myblue}{rgb}{0,0,0.75}
\definecolor{mymagenta}{cmyk}{0,1,0,0.12}

\begin{document}

\title{Long-term laser frequency stabilization using  fiber interferometers} 



\author{Jia Kong}
\email[Corresponding author]{jia.kong@icfo.es}
\affiliation{ICFO -- Institut de Ciencies Fotoniques,
Mediterranean Technology Park, 08860 Castelldefels, Barcelona,
Spain}
\affiliation{Quantum Institute for Light and Atoms, Department of Physics,
State Key Laboratory of Precision Spectroscopy, East China Normal
University, Shanghai 200062, China}
\author{Vito Giovanni Lucivero}
\affiliation{ICFO -- Institut de Ciencies Fotoniques,
Mediterranean Technology Park, 08860 Castelldefels, Barcelona,
Spain}
\author{Ricardo Jim\'enez-Mart\'inez}
\affiliation{ICFO -- Institut de Ciencies Fotoniques,
Mediterranean Technology Park, 08860 Castelldefels, Barcelona,
Spain}
\author{Morgan W. Mitchell}
\affiliation{ICFO -- Institut de Ciencies Fotoniques,
Mediterranean Technology Park, 08860 Castelldefels, Barcelona, Spain}
\affiliation{ICREA -- Instituci\'o Catalana de Recerca i Estudis Avan\c{c}ats, 08015 Barcelona, Spain}

\date{\today}

\begin{abstract}
We report long-term laser frequency stabilization using only the
target laser and a pair of 5 m fiber interferometers, one as a
frequency reference and the second as a sensitive thermometer
to stabilize the frequency reference. When used to
stabilize a distributed feedback laser at 795 nm, the frequency
Allan deviation at 1000~s drops from $5.6\times10^{-8}$ to
$6.9\times10^{-10}$.
The performance equals that of an offset lock employing a
second, atom-stabilized laser in the temperature control.
\end{abstract}

\maketitle

\section{Introduction}
Laser linewidth and frequency stability are critical in
laser spectroscopy and its many applications. Short-term linewidth
can be narrowed by active feedback from a fast frequency
discriminator such as an optical cavity \cite{DreverAPB1983} or an
interferometer \cite{ChenAO1989}, and is limited by the speed of
the feedback and the noise of the discriminator signal
\cite{KesslerNPhot2012}. Long-term stability above hundreds
milliseconds \cite{KesslerNPhot2012,RossiRSI2002}, in contrast,
requires stabilization to an absolute reference and is limited by
slow changes in the reference and in the feedback system. As references,
atomic and molecular lines are very stable but give low signal to
noise ratios and a limited selection of frequencies
\cite{Demtroeder2003,TsuchidaJJAP1982}. For this reason, if
long-term stability at a frequency far from atomic and molecular
lines is needed, linewidth narrowing is often combined with a
¡°transfer lock¡± in which a first laser is stabilized to an atomic
or molecular transition, a discriminator is stabilized to this
laser, and a second laser is stabilized to the discriminator,
possibly at a different wavelength \cite{RossiRSI2002}.

The complexity of this approach can in principle be avoided if the
frequency discriminator itself provides a stable reference. Here
we demonstrate stabilization of a distributed-feedback
(DFB) diode laser to two unbalanced Mach-Zehnder interferometers
(MZIs), one used to stabilize the temperature of the other,
interrogated with the same target laser. The system derives its
stability from the material properties of silica fiber and a
metal, in our case an aluminum alloy. We observe the same
long-term stability as using two independent lasers for
MZIs, e.g. $6.9 \times 10^{-10}$ at 1000 s. Although much less
stable than the best optical cavities
\cite{KesslerNPhot2012}, our setup provides long-term stability with lower cost and complexity. An application requiring this level of stability is quantum-enhanced magnetometry \cite{WolfgrammPRL2010,LuciveroRSI2014}, which will also require many-GHz detunings in the spin-exchange-relaxation-free regime \cite{Jimenez-MartinezJOSAB2012,Jimenez-MartinezRSI2014}.

Unbalanced fiber interferometers have recently emerged as
suitable references to sense and stabilize laser frequency. The
interferometer phase $\phi$ is very sensitive to laser frequency
$f$ due to the large physical path difference $L$:
\begin{equation}
\phi=\frac{2\pi n}{c}f L,
\end{equation}
where $n$ is the refractive index of the fiber and $c$ is the
speed of light. Prior work includes MZI stabilization of a
helium-neon laser to 5 kHz linewidth \cite{ChenAO1989} over a time
scale of 1 s, stabilization erbium-doped fiber
distributed-feedback lasers
(EDFLs)\cite{CranchOL2002,TakahashiJP2008,KefelianOL2009}, e.g. to
8 Hz linewidth \cite{KefelianOL2009} over 1 s using a 2 km
path-imbalanced Michelson interferometer (MI). The linewidth of
DFB diode lasers \cite{ClicheSPIE2006,LeeRSI2011} has
shown large narrowing factors, e.g. from 370 kHz to 18 Hz over 1
ms using a MI \cite{ClicheSPIE2006} and from 3 MHz to 15 kHz using a MZI \cite{LeeRSI2011}.
50 Hz peak-to-peak non-linearity frequency error was also achieved in an agile laser with high sweep linearity \cite{JiangOE2010}.

Long-term stability has been little studied with unbalanced
interferometers and the above works only achieved short-term
linewidth reduction with long-term stability affected by
temperature fluctuations. For example, in \cite{LeeRSI2011} there
is no stability improvement over times longer than 64~s.
To achieve long-term stability, Chiodo \emph{et al.} \cite{ChiodoAO2013} used a high-precision electronic temperature controller with well-designed physics package \cite{BoudotRSI2005} to control the fiber's temperature. Without high-precision electronics, Wang \emph{et al.}
\cite{WangAO2014} implemented a transfer lock using an atomic
reference to control the temperature of a 2 m
path-imbalanced Young's interferometer, and stabilized an external
cavity diode laser to $10^{-8}$ over 10-4000 s.

\begin{figure*}[t]
\centerline{\includegraphics[width=150mm]{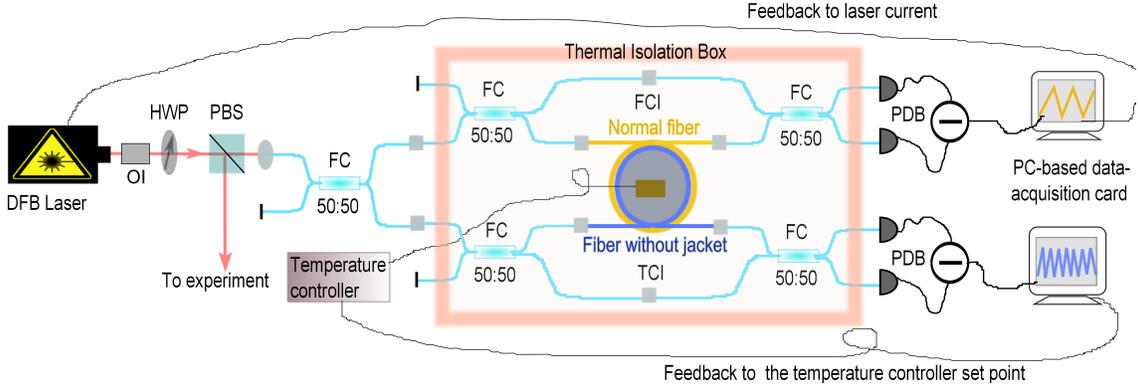}}
\caption{\label{Install} Schematic of laser frequency locking
system. Using a half-wave plate (HWP) and a polarizing beam
splitter (PBS), a few 10s of $\mu$W of power from the laser to be
stabilized, a DFB (EYP-DFB 795) with wavelength 795 nm followed by an optical isolator (OI), is injected into a fiber
system containing two unequal-path Mach-Zehnder interferometers.
Each interferometer is composed of two 50:50 fiber couplers (FC)
and an additional 5 m fiber to imbalance the paths. In the
frequency control interferometer (FCI) the extra fiber is a
jacketed single mode fiber, while for the temperature control
interferometer (TCI), it is a polarization maintaining fiber
without jacket. The fibers are wound around a 10 cm diameter
aluminum cylinder fitted with a resistive heater and a resistive
temperature sensor. In each interferometer the output powers are
collected on a Thorlabs balanced amplified photodetector
(PDB450C), providing signals that are used for feedback, either to
the laser current, or to the set-point of a temperature controller
from Wavelength Electronics (model HTC1500) by a digital
controller (PC-based data-acquisition card). The interferometers
are enclosed in a insulation layer of extruded polystyrene foam.}
\end{figure*}

\section{Operating principle}
In our scheme, one interferometer, the ``frequency control
interferometer'' (FCI), is used to sense and stabilize the laser
frequency, while a second, ``temperature control interferometer''
(TCI) is used to sense and eliminate fluctuations in the
temperature of the thermal reservoir to which both are attached.
As both interferometers are interrogated using the same
target laser, this scheme  provides a simpler alternative to
approaches relying on a second reference laser \cite{RossiRSI2002,
WangAO2014}. The FCI and TCI are constructed to have the same
frequency response
\begin{equation}
\frac{d\phi}{df}=\frac{2\pi n L}{c} = 0.1522 \,\, \frac{\rm
rad}{\rm MHz}, \label{F}
\end{equation}
where $n=1.454$ \cite{LevitonSPIE2006} and $L=5$ m (equal for the
two interferometers). Larger $L$ would give better frequency
discrimination, but a larger device.

The TCI is designed to have a much stronger temperature response
than the FCI, which can be achieved by reducing the temperature
sensitivity of the latter by using a special fiber
\cite{DanguiOE2005}. Here we instead increase the temperature
sensitivity of the TCI by using a jacketless fiber, tightly wound
on an aluminum cylinder, as the extra length in the TCI. The
coefficient of thermal expansion (CTE) $\alpha = L^{-1}dL/dT$ of
aluminum, $\alpha_{\rm Al}=23\times10^{-6}$ K$^{-1}$, is 40 times
larger than the CTE of silica fiber $\alpha_{\rm
SiO_2}=0.5\times10^{-6}$ K$^{-1}$, while the two materials have
very similar Young's moduli, so the fiber is stretched by the
aluminum cylinder as the temperature $T$ rises, giving $dL_{\rm
TCI} / dT > dL_{\rm FCI} / dT$.

In the TCI, phase change with temperature arises from the
interaction of the thermo-optic coefficient $dn/dT$, the CTE, and
the elasto-optic coefficient $dn/dL$
\begin{eqnarray}
\frac{d\phi _{\rm TCI}}{dT}=\frac{2\pi L}{\lambda} \frac{dn}{dT}+
\frac{d\phi_{\rm TCI}}{dL_{\rm TCI}}\frac{dL_{\rm TCI}}{dT},
\label{TCI1}
\end{eqnarray}
where $\lambda=c/f$. The first part of Eq.~(\ref{TCI1}) describes
the thermo-optic effect, while  the second part combines the other
two effects. In contrast the FCI is not stretched, so there is no
elasto-optic effect and the phase change with temperature is
\begin{eqnarray}
\frac{d\phi_{FCI}}{dT}=\frac{2\pi L}{\lambda}
\frac{dn}{dT}+\frac{2\pi n}{\lambda} \frac{dL_{FCI}}{dT}.
\label{FCI1}
\end{eqnarray}
With $\lambda=795$ nm, $dn/dT=9.2\times10^{-6}$ K$^{-1}$,
$d\phi_{\rm TCI}/dL_{\rm TCI}=9.14\times10^{6}$ rad/m
 \cite{Silva-LopezOL2005}, $L^{-1} dL_{\rm
TCI}/dT=\alpha_{Al}$, and $L^{-1} dL_{\rm FCI}/dT=\alpha_{SiO_2}$
we find $d\phi _{\rm TCI}/dT= 1414$ rad/K and $d\phi _{\rm
FCI}/dT= 392$ rad/K and thus the TCI is 3.6 times more sensitive
to temperature than the FCI.

\section{Experimental realization}
The setup is shown and described in FIG. ~\ref{Install}. To
stabilize the frequency of a DFB laser, we use MZIs with
differential detection, which allows rejecting common-mode
intensity fluctuations, a two-fold increase in slope because the
outputs are anti-phase, and in principle, shot-noise limited
performance. In the TCI, the additional 5 m fiber has no jacket
but retains a 60 $\mu$m acrylate coating applied during
manufacture. This fiber is tightly wrapped by hand around the
aluminum cylinder while at room temperature. The expansion to
reach the 60 degree Celsius operating point exceeds the coating
thickness and guarantees the fiber is always under tension.
In this case, a polarization maintaining (PM) fiber is used
to prevent stress-induced polarization fluctuation. The 5 m fiber of the FCI,
which has a 328 $\mu$m jacket, is wrapped on the same aluminum
cylinder. This guarantees good thermal contact of the two fibers
with a single thermal reservoir, but the FCI does not stretch
significantly.

\begin{figure}[t]
\includegraphics[width=90mm]{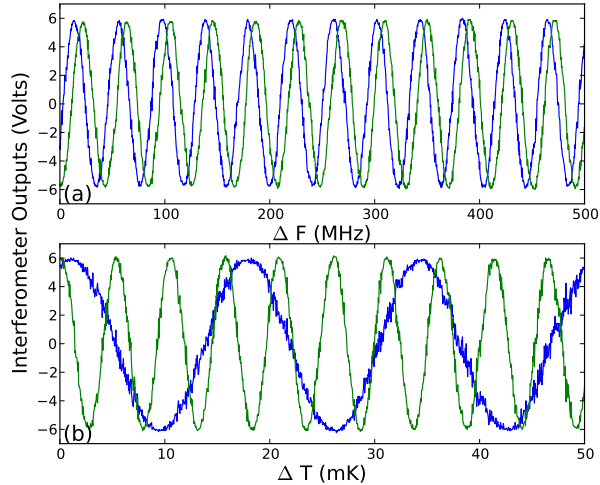}
\caption{\label{interference} Interferometers' phase response to
temperature and frequency. (a) Differential outputs of the FCI
(blue) and TCI (green) as laser frequency is scanned by
$\Delta F$.} (b) Same signals as interferometers' temperature
is scanned by $\Delta T$. While the frequency response is
matched, the TCI is about three times more sensitive to
temperature changes.
\end{figure}

FIG. ~\ref{interference} shows the two interferometer's
responses to frequency and temperature scans. As there is a
minor length difference between FCI and TCI, their responses to
frequency are not perfectly matched. This does not strongly affect the stabilization, which maintains a unique frequency/temperature combination provided the interferometers' sensitivity to temperature and to frequency are different.  We calibrate the frequency
response against a $^{87}$Rb absorption spectrum to find 0.1544
rad/MHz (TCI), 0.1540 rad/MHz (FCI), which are close to the 0.1522
rad/MHz expected from Eq.~(\ref{F}). We calibrate the temperature
response against a thermistor and find 1211 rad/K (TCI) and 400
rad/K (FCI), which agree reasonably well with the values found
above.

The output of the FCI can be fed back to the laser current to
stabilize the laser frequency. The output of the TCI can similarly
be fed back to the set point of the temperature controller to
stabilize the temperature of the aluminum cylinder. Both controls
are realized with a 100 kHz bandwidth data acquisition card.
Without the feedback from TCI, the thermal gain (ratio of laboratory fluctuations to system fluctuations) of the system is about 400, and it improves by a factor of 3 when the TCI is used as temperature probe to further stabilize the set point of the temperature controller.

\section{Results and discussions}
\begin{figure}[t]
\includegraphics[width=90mm]{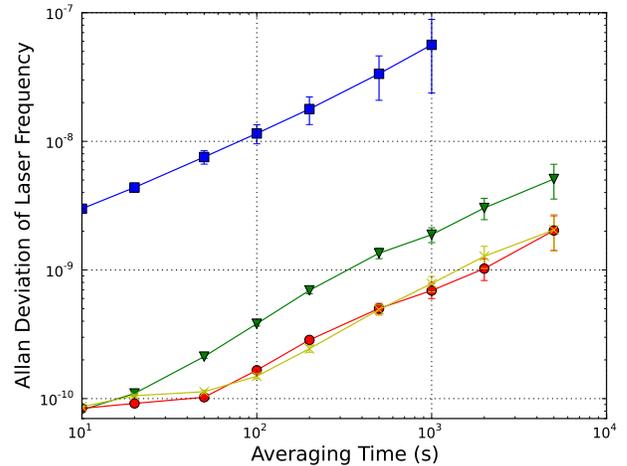}
\caption{\label{Allan} Allan deviation of DFB laser frequency
under various control scenarios. Free running laser (blue
squares), frequency stabilization by FCI feedback to laser current
(green triangles), temperature stabilization by TCI feedback to
temperature controller, in addition to frequency stabilization
(red circles), and temperature stabilization by TCI feedback, but
with the TCI injected by a separate laser frequency stabilized by
saturated-absorption spectroscopy (yellow stars). Error bars are
obtained by dividing by the square root of the number of samples
in each averaging time interval \cite{Howe}. }
\end{figure}

To monitor the frequency fluctuations of the target laser, we interfered
the laser output against a second laser stabilized by saturated
absorption spectroscopy to the D$_1$ line of Rb. This
reference laser had a stability, measured by beating against a duplicate laser,
of $\le 8 \times 10^{-11}$ at  $1000$ s and a $t^{1/2}$ scaling, an
order of magnitude better
than the lasers under test. The resulting
beat note was recorded on a spectrum analyzer and the computed
centroid of the spectrum was taken as the current frequency.
We collected the frequency every 10 seconds over 15 hours
of total acquisition, and compute the Allan deviation
\cite{AllanIEEE1966} for various control scenarios, shown and
described in FIG. ~\ref{Allan}.

As expected, stabilization to the FCI provides
better stability, by about a factor of 30 for all time scales measured.
For times above about 1 minute, the TCI provides a further
improvement by about a factor of 2.5 which is larger in longer
time scale, giving a relative frequency stability (Allan deviation) of $8.5 \times 10^{-11}$  at 10~s and
$2.0 \times 10^{-9}$ at 5000~s.  Multiplying these numbers by the laser central frequency
 we find 32 kHz at 10 s and 754 kHz at 5000 s. The long-time scaling is $t^{1/2}$, characteristic of a frequency random walk.
To show how the different temperature response of
the FCI and TCI can help a single laser to distinguish between
variations in frequency and in temperature, we used a second laser,
stabilized to the saturated absorption spectrum of Rb-85 D$_1$
line as the input to the TCI for temperature stabilization. The
resulting Allan deviation is indistinguishable from that observed
by self-stabilization of the DFB laser, demonstrating that
our measurement is not limited by using the target laser for
temperature stabilization. This makes our technique
competitive with the transfer lock, while being inexpensive,
compact and flexible. Moreover, with this technique, the laser can
be locked far from an atomic or molecular frequency reference by
counting the number of interference fringes.

We have concentrated on improving long-term frequency
stability, as short-term stabilization with fiber interferometers
has been well studied \cite{LeeRSI2011}. A single fiber interferometer
can provide both short-term and long-term stability, using a fast feedback controller with good long-term stability. Using a 5 m path-imbalanced fiber interferometer and high-bandwidth feedback,
they narrowed the linewidth of a DFB laser from 3 MHz to 15 kHz.
Combined with our self-referencing method, the laser linewidth can
be reduced for both short  and long time scales.

Atomospheric pressure, vibration and polarization
fluctuations are other factors that can limit this locking
performance, so vacuum tank and vibration isolation can be used
for further improvement \cite{ChenAO1989,KefelianOL2009}.
Temperature inhomogeneity may still exist, which could be reduced
by interleaving the two fibers on their mutual support.

\section{Conclusion}
In conclusion, we have described a flexible long-term laser
frequency stabilization method using two interferometers with very
different temperature coefficients.  Using only a single laser, we
can lock to frequencies not corresponding to any atomic or
molecular line. We observe an Allan deviation of
$6.9\times10^{-10}$ at 1000 s, an improvement by a factor of 81
relative to the laser with electronic temperature and current
stabilization. The laser stability can achieve $8.5 \times 10^{-11}$  at 10~s and
$2.0 \times 10^{-9}$ at 5000~s. The method is compatible with short-term
linewidth narrowing and with integrated interferometers, promising
a small, robust, cheap and flexible DFB laser with both short-term
and long-term frequency stability.

\section{Acknowledgments}
{We thank Federica Beduini for technical help and useful
discussions, Thomas Vanderbruggen for advice on the data analysis
and Joanna Zieli\'{n}ska for the loan of a balanced photodetector.
This work was supported by the Spanish MINECO project MAGO (Ref.
FIS2011-23520), European Research Council project AQUMET and
Funaci\'{o} Privada CELLEX. J. K. is supported by the China
Scholarship Council project and the Fostering Project for National
Top Hundred Doctoral Dissertations (No. PY2013006). }

\bibliography{LaserStabilization}

\end{document}